\documentclass[twocolumn,showpacs,preprintnumbers,amsmath,amssymb,floatfix]{revtex4-1}

\usepackage{graphicx}
\usepackage{dcolumn}
\usepackage{bm}
\usepackage{latexsym}
\usepackage[hang,small,bf]{caption}
\usepackage[subrefformat=parens]{subcaption}
\usepackage{here}
\usepackage{multirow}

\usepackage{float}
\def\Tr{\text{Tr}} 

\newcommand{\beqn}{\begin{eqnarray}}
\newcommand{\beq}{\begin{equation}}
\newcommand{\eeqn}{\end{eqnarray}}
\newcommand{\eeq}{\end{equation}}

\newcommand{\tr}{\mathop{\rm Tr}}

\DeclareMathOperator*{\TrTr}{Tr}

\usepackage{braket}
\newcommand{\RCNP}{\affiliation{Research Center for Nuclear Physics (RCNP), Osaka University, Osaka 567-0047, Japan}}
\newcommand{\PhDKochi}{\affiliation{Graduate School of Integrated Arts and Sciences, Kochi University, Kochi 780-8520, Japan}}
\newcommand{\Kochi}{\affiliation{Library and Information Technology, Kochi University, Kochi 780-8520, Japan}}

\begin{document}
\title{New Abelian-like monopoles and the dual Meissner effect}
\author{Atsuki Hiraguchi}
\email[e-mail:]{b18d6a04@s.kochi-u.ac.jp}
\PhDKochi
\author{Katsuya Ishiguro}
\Kochi
\author{Tsuneo Suzuki}
\RCNP

\date{\today}

\begin{abstract}
Violation of non-Abelian Bianchi identity can be regarded as $N^2-1$ Abelian-like monopole currents in the continuum SU(N) QCD. Three Abelian-like monopoles, when defined in SU(2) gluodynamics on the lattice \`{a} la DeGrand-Toussaint, are shown to have the continuum limit with respect to the color-invariant monopole density and the effective monopole action. Since each Abelian-like monopole is not gauge invariant, we have introduced various partial gauge fixing for the purpose of reducing lattice artifact monopoles in the thermalized vacuum. Here we investigate Abelian and monopole dominances and the Abelian dual Meissner effects adopting the same gauges like the maximal center gauge (MCG) in comparison with the maximal Abelian gauge (MAG). Abelian and monopole contributions to the string tension in these gauges are observed to be a little smaller than the non-Abelian string tension. However, we find that the monopole dominance is improved well when use is made of  the block-spin transformations with respect to Abelian-like monopoles. We find each electric field is squeezed by the corresponding colored Abelian-like monopole in such gauges and the Abelian dual Meissner effect is observed independently for each color.  Moreover, we confirm the  dual Amp$\grave{\mathrm{e}}$re's law in these new gauges as well as in MAG. The SU(2) vacuum is shown to be near the border between the type 1 and type 2 dual superconductors. The penetration length is almost equal for the four gauge fixings and the vacuum type in MCG is almost the same value as the previous results. These results are consistent with the previous results suggesting the continuum limit and the gauge-independence of Abelian monopoles. 
\end{abstract}

\maketitle
\section{introduction}
The mechanism of color confinement is still unknown in quantum 
chromodynamics (QCD) \cite{CMI:2000mp}.  

As a picture of color confinement, 't~Hooft~\cite{tHooft:1975pu} and Mandelstam~\cite{Mandelstam:1974pi} conjectured that the QCD vacuum is a dual superconducting state. An interesting idea to realize this conjecture is to project QCD to the Abelian maximal torus group by a partial (but singular) gauge fixing~\cite{tHooft:1981ht}. After the Abelian projection, color magnetic monopoles appear as a topological current. The dual Meissner effect is caused by condensation of monopoles. Numerically, Abelian monopole dominance is observed clearly in the maximal Abelian gauge (MAG) fixing~\cite{Kronfeld:1987ri,Kronfeld:1987vd,Suzuki:2002}. Similar results are found also in various local unitary gauges~\cite{Sekido:2007mp}. However, there are infinite ways of such a partial gauge fixing. It is not at all
clear if the lattice results obtained in a partial gauge fixing like MAG are gauge independent.  In the works~\cite{Suzuki:2007jp,Suzuki:2009xy}, Abelian monopole dominance and the dual Meissner effect are found to exist even without adopting any gauge fixing. By making use of a huge number of thermalized vacua 
with additional random gauge transformations, they found that the string tension from the monopole Polyakov loop correlations is identical to that of the gauge-invariant non-Abelian static potential. There exists also the Abelian dual Meissner effect. The vacuum type of pure SU(2) gauge theory was found to be near the border between type 1 and type 2 dual superconductors. Although the results are interesting, the physical meaning of such gauge-variant quantities without gauge fixing was not clear at all in the continuum limit of QCD. 

Recently, it was shown in the continuum limit that the violation of the non-Abelian Bianchi identities (VNABI) $J_{\mu}$ is equal to Abelian-like monopole currents $k_{\mu}$ defined by the violation of the Abelian-like Bianchi identities \cite{Suzuki:2014wya,Suzuki:2017lco}. Although VNABI is a gauge-variant adjoint operator satisfying the covariant conservation rule, it gives us, at the same time, the Abelian-like current conservation rules. There are $N^2-1$ conserved Abelian-like magnetic charges in the case of color SU(N). We can define lattice Abelian-like monopoles following DeGrand-Toussaint \cite{DeGrand:1980eq}. They are just equal to the gauge-variant lattice Abelian-like monopoles studied previously in the works \cite{Suzuki:2007jp,Suzuki:2009xy}.  

 It is interesting to study the continuum limit of the lattice  Abelian-like monopole density. But the technique adopted in Ref.\cite{Suzuki:2007jp,Suzuki:2009xy} can not be applied since such a quantity as monopole densities is always positive definite.  Therefore, various techniques were introduced for the thermal vacua \cite{Suzuki:2017lco}, which are contaminated by lattice artifacts originally. Gauge fixing and the monopole block-spin transformation \cite{Shiba:2005} are two main techniques extracting physical quantities from the lattice vacuum. With such gauge fixings, they adopted three global color invariant gauges: the maximal Center gauge (MCG) \cite{DelDebbio:1996mh,DelDebbio:1998uu}, the direct Laplacian center gauge (DLCG) \cite{Faber:2001zs}, and the maximal Abelian Wilson loop gauge (MAWL) \cite{Suzuki:1996ax}, as well as the global color-variant MAG \cite{Kronfeld:1987ri,Kronfeld:1987vd,Suzuki:2002} with additional {U(1)} Landau gauge fixing (MAU1). Then convincing scaling behaviors are seen when the density $\rho(a(\beta),n)$ is plotted versus the lattice spacing of the blocked lattice $b=na(\beta)$, where $a(\beta)$ is the lattice distance at the coupling $\beta$. A single universal curve $\rho(b)$ is found from $n=1$ up to $n=12$ for all four gauges adopted, which suggests that $\rho(a(\beta),n)$ is a function of $b=na(\beta)$ alone and gauge independent. Since the continuum limit is realized as $n\to\infty$, scaling means that the lattice definition of 
Abelian-like monopoles has a continuum limit. Afterwards, one of the present authors (T.S.) found that coupling constants of the effective monopole action derived from the inverse  Monte Carlo method \cite{swendsen,Shiba:2005} show also a universal scaling behavior \cite{Suzuki:2017zdh} for the above four gauges. 

It is the purpose of this work first to investigate whether the Abelian monopole dominance and the Abelian dual Meissner effect, which are observed in the MAG \cite{Suzuki:2002}, are seen also in the above global color invariant gauges (MCG, DLCG, and MAWL) with a reasonable number of field configurations. Since VNABI are gauge variant, finding various gauge-fixing methods reducing lattice artifact monopoles without destroying physical monopole effects is very important for extracting any physical quantity concerning Abelian-like monopoles. Hereafter, the authors call such gauges as smooth in this work. Secondly, it is interesting to check global-color independence of the Abelian dual Meissner effect when we introduce a single color external source in the vacuum. Such a study could not be done in practice at the present stage
without any gauge fixing as in Ref.\cite{Suzuki:2009xy}. Hence, we adopt here the above global color invariant gauges smoothing the vacuum. Furtheremore, we use the block-spin transformation of the monopole current in comparing the monopole contribution to confinement in the MCG with that in the MAG.

\section{Method}
\subsection{The violation of non-Abelian Bianchi identities}
If gauge fields have a line singularity in the continuum QCD, then the non-Abelian Bianchi identity is violated. 
The VNABI is found to be equivalent to that of Abelian-like Bianchi identity \cite{Suzuki:2014wya,Suzuki:2017lco}. Namely VNABI is regarded as eight Abelian-like monopoles in the continuum SU(3) QCD. Using a covariant derivative $D_{\mu}=\partial_{\mu} - igA_{\mu}$, we get the following  commutation relation:
\begin{align}
  [D_{\mu},D_{\nu}] = -igG_{\mu\nu} + [\partial_{\mu},\partial_{\nu}],
\end{align}
where  $G_{\mu\nu}$ is a non-Abelian field strength. The second commutator can not be discarded when a line singularity exists. The Jacobi identities, 
 \begin{align}
  \epsilon_{\mu \nu \rho \sigma}[D_{\nu},[D_{\rho},D_{\sigma}]]=0 ,
 \end{align}
lead us to the following relation:
\begin{align}
 D_{\nu}  G^{*}_{\mu \nu} = \partial_{\nu} f^{*}_{\mu \nu} = k_{\mu}, \label{VNABI}
\end{align}
where $f_{\mu \nu}$ is defined as $f_{\mu \nu} = \partial_{\mu}A_{\nu} - \partial_{\nu} A_{\mu} = (\partial_{\mu} A_{\nu}^a - \partial_{\nu} A_{\mu}^a)\lambda^a/2$.
In the case of SU(3), $\lambda^a$ are the Gell-Mann matrices. 
Relation (\ref{VNABI}) means that the VNABI is equivalent to eight Abelian-like magnetic monopole currents in SU(3).  In the case of SU(2), VNABI is equivalent to three Abelian-like magnetic monopole currents.
\subsection{Abelian-like monopoles on a lattice}
The direct definition of VNABI on lattice is very difficult. Hence, we adopt defining Abelian-like monopoles on a lattice following Ref.\cite{DeGrand:1980eq} and study the continuum limit of them since VNABI is equivalent to Abelian-like monopoles in the continuum limit. 

We consider here also SU(2) gluodynamics for simplicity. SU(2) link variables are  
\begin{align}
 U_{\mu}(s) = U^0_{\mu}(s) + i \sigma^a U^a_{\mu}(s), 
\end{align}
where $\sigma^a$ are the Pauli matrices, $a=1,2,3$ are color indices and $U^0_{\mu}(s), U^a_{\mu}(s)$ are real coefficients. We explain how to define Abelian-like monopoles in SU(2) gauge theory below. \\
\ \ First, Abelian-like gauge fields $\theta_{\mu}^a(s)$ are derived to get the maximum overlap with an original non-Abelian link variable, namely in such a way as the following quantity is maximized:
\begin{align}
R_1 = \sum_{s,\mu} \mathrm{Re}\Tr[e^{i\theta_{\mu}^1(s)\sigma^1} U_{\mu}^{\dagger}(s)] ,
\end{align}
where only the case for color=1 is written as an example. Then we get
\begin{align}
\theta_{\mu}^a(s) &= \mathrm{arctan}\left(\frac{U^a_{\mu}(s)}{U^0_{\mu}(s)}\right) \  (|\theta^a_{\mu}(s)|<\pi ). \label{Abelian}
\end{align}
This is equal to the definition as adopted in previous works \cite{Suzuki:2007jp,Suzuki:2009xy}.

We now define three monopole currents following Ref.\cite{DeGrand:1980eq}:
\begin{align} 
k_{\nu}^a(s) &= \frac{1}{4\pi}\epsilon_{\mu\nu\rho\sigma}\partial_{\mu} \bar{\theta}^a_{\rho \sigma}(s+\hat{\nu}), \\
&\theta^a_{\mu \nu}(s) = \partial_{\mu} \theta^a_{\nu}(s) - \partial_{\nu}\theta^a_{\mu}(s), \notag \\
&\bar{\theta}^a_{\mu\nu}(s) = \theta^a_{\mu \nu}(s) - 2\pi n^a_{\mu \nu}(s), \notag
\end{align}
where $\theta^a_{\mu \nu}(s)$ is an Abelian-like field strength, $\bar{\theta}^a_{\mu\nu} \in [-\pi,\pi]$, and $n^a_{\mu \nu}(s)$ is an antisymmetric tensor. Note that $n^a_{\mu \nu}(s)$ takes integer values \{-2,-1,0,1,2\}. It can be interpreted as the number of Dirac strings. We found these monopole currents have a continuum limit, studying the monopole density and the effective monopole action on the lattice with the aid of a block-spin transformation of monopoles \cite{Suzuki:2017lco,Suzuki:2017zdh}. 

\subsection{Smooth gauge fixings}
We adopt gauge-fixing techniques smoothing the vacuum as in Ref.\cite{Suzuki:2017lco}. The gauge-fixing methods adopted here reduce lattice artifact monopoles well without destroying infrared long monopoles.
\begin{enumerate}
  \item \textbf{MCG}. The first is the maximal center gauge \cite{DelDebbio:1996mh,DelDebbio:1998uu}, which is usually discussed in the framework of the center vortex idea. We adopt the so-called direct maximal center gauge, which requires maximization of the quantity
\begin{eqnarray}
R=\sum_{s,\mu}(\tr U_{\mu}(s))^2 \label{eq:MCG} ,
\end{eqnarray}
with respect to local gauge transformations.  The condition (\ref{eq:MCG}) fixes the gauge up to
$Z(2)$ gauge transformation. 
\item \textbf{DLCG}. The second is the Laplacian center gauge \cite{Faber:2001zs}, which is also discussed in connection with the  center vortex idea. 
\item \textbf{MAWL}. Another is the maximal Abelian Wilson loop gauge, in which
\begin{eqnarray}
R&=&\sum_{s,\mu\neq\nu}\sum_a(\cos(\theta^a_{\mu\nu}(s))) , \label{SAWL}
\end{eqnarray}
is maximized  \cite{Suzuki:1996ax}. 
Since $\cos(\theta^a_{\mu\nu}(s))$ are $1\times 1$ Abelian Wilson loops, the gauge is called the maximal Abelian Wilson loop gauge.
\item \textbf{MAU1}. The fourth is the combination of MAG and the U(1) Landau gauge \cite{Kronfeld:1987ri,Kronfeld:1987vd}. Namely, we first perform maximal Abelian gauge fixing and then, with respect to what remains, U(1) symmetry Landau gauge fixing is done. This case breaks the global SU(2) color symmetry contrary to the previous three cases (MCG, DLCG, and MAWL), but we consider this case since the vacuum is smoothed fairly well. The MAG is the gauge which maximizes
\begin{equation}
  R=\sum_{s,\mu}{\rm Tr}\Big(\sigma_3 U_{\mu}(s)
        \sigma_3 U^{\dagger}_{\mu}(s)\Big) , \label{R}
\end{equation}
with respect to local gauge transformations. Then there remains a U(1) symmetry to which the Landau gauge fixing is applied, i.e.,
$ \sum_{s,\mu} (\cos \theta^3_{\mu}(s))$   is maximized \cite{Bali:1996dm}.
\end{enumerate}

\subsection{Simulation details}
 In most cases we adopt the tadpole improved action in pure SU(2) gauge theory: 
\begin{align}
 S = \beta \{ \sum_{pl} S_{pl} - \frac{1}{20u_{0}^2} \sum_{rt} S_{rt} \} , \label{tad}
\end{align}
where $S_{pl}$ and $S_{rt}$ denote plaquette and $1\times2$ rectangular loop terms in the action, 
\begin{align}
 S_{pl,rt} = \frac{1}{2} \TrTr(1-U_{pl.rt}), 
\end{align}
the parameter $u_0$ is the input tadpole improvement factor taken here equal to the fourth root of the average plaquette $P=\braket{\frac{1}{2}trU_{pl}}$. In our simulations we do not include one-loop corrections to the coefficients for the sake of simplicity. The lattices adopted are $48^4$ for $\beta=3.0 \sim 3.9$ and $24^4$ for $\beta=3.0 \sim 3.9$. In the case of the tadpole improved action, we adopt the same vacuum ensembles generated and used in the previous research \cite{Suzuki:2017lco}.

\section{Results}
\begin{table}[h] 
\caption{The string tension at $\beta=3.5$ on the $24^4$ lattice}\label{fourGauge-sigma}
\begin{center}
\begin{tabular}{|c|c|c|c|c|c|} \hline
 \multicolumn{2}{|c|}{} &  \multicolumn{1}{|c|}{$\sqrt{\sigma a^2}$}&  \multicolumn{1}{|c|}{$\sigma_{A}/\sigma_{NA}$}  &  \multicolumn{1}{|c|}{$\sigma_{mon}/\sigma_{NA}$}  &  \multicolumn{1}{|c|}{$\sigma_{ph}/\sigma_{NA}$}\\ \hline
 \multirow{4}{*}{MCG}&$V_{NA}$&$0.1555(6)$ &\multirow{4}{*}{0.8149} &\multirow{4}{*}{0.7053}& \multirow{4}{*}{0.3709} \\
                              &$V_{A}$&0.1267(7)& && \\
                              &$V_{mon}$&0.1096(3)&&& \\ 
                              &$V_{ph}$&0.0576(2)&&& \\ 
\hline
\multirow{4}{*}{DLCG}&$V_{NA}$&$0.1555(6)$&\multirow{4}{*}{0.8316}&\multirow{4}{*}{0.708}&\multirow{4}{*}{0.3605} \\
                              &$V_{A}$&0.1293(6)&&& \\
                              &$V_{mon}$&0.1100(5)&&& \\
                              &$V_{ph}$&0.0560(2)&&& \\ 
\hline
\multirow{4}{*}{MAWL}&$V_{NA}$&$0.1555(6)$&\multirow{4}{*}{0.8151}&\multirow{4}{*}{0.7066}&\multirow{4}{*}{0.3693} \\
                               &$V_{A}$&0.1267(7)&&& \\
                               &$V_{mon}$&0.1098(3)&&& \\
                               &$V_{ph}$&0.0574(3)&&& \\ 
\hline
\multirow{4}{*}{MAU1} &$V_{NA}$&$0.1555(6)$&\multirow{4}{*}{0.8778}&\multirow{4}{*}{0.722}&\multirow{4}{*}{0.4114} \\
                               &$V_{A}$&0.136(1)&&&\\
                               &$V_{mon}$&0.1122(2)&&& \\
                               &$V_{ph}$&0.0639(9)&&& \\ 
\hline
\end{tabular}
\end{center}
\end{table}
\begin{table}[h] 
\caption{The string tension in the MCG on the $48^4$ lattice}\label{MCG-sigma}
\begin{center}
\begin{tabular}{|c|c|c|c|c|c|} \hline
 \multicolumn{2}{|c|}{} &  \multicolumn{1}{|c|}{$\sqrt{\sigma a^2}$}&  \multicolumn{1}{|c|}{$\sigma_{A}/\sigma_{NA}$}  &  \multicolumn{1}{|c|}{$\sigma_{mon}/\sigma_{NA}$} &  \multicolumn{1}{|c|}{$\sigma_{ph}/\sigma_{NA}$}\\ \hline
\multirow{4}{*}{$\beta = 3.0$}&$V_{NA}$&$0.3728(4)$&\multirow{4}{*}{0.8923}&\multirow{4}{*}{0.7807}&\multirow{4}{*}{0.1794} \\
                                         &$V_{A}$&0.3326(3)&&& \\
                                         &$V_{mon}$&0.2910(1)&&& \\
                                         &$V_{ph}$&0.066(1)&&& \\ 
\hline
\multirow{4}{*}{$\beta = 3.2$}&$V_{NA}$&$0.2630(4)$&\multirow{4}{*}{0.8541}&\multirow{4}{*}{0.7576}&\multirow{4}{*}{0.2110} \\
                                         &$V_{A}$&0.2246(3)&&& \\
                                         &$V_{mon}$&0.1992(2)&&&  \\
                                         &$V_{ph}$&0.0554(4)&&& \\ 
\hline
\multirow{4}{*}{$\beta = 3.5$}&$V_{NA}$&$0.1546(3)$&\multirow{4}{*}{0.8525}&\multirow{4}{*}{0.7604}&\multirow{4}{*}{0.2701} \\
                                        &$V_{A}$&0.1317(3)&&&\\
                                        &$V_{mon}$&0.1175(4)&&& \\
                                        &$V_{ph}$&0.0417(1)&&& \\
\hline
\multirow{4}{*}{$\beta = 3.9$}&$V_{NA}$&$0.0829(2)$&\multirow{4}{*}{0.9283}&\multirow{4}{*}{0.77}&\multirow{4}{*}{0.3841} \\
                                         &$V_{A}$&0.0769(4)&&&\\
                                         &$V_{mon}$&0.0638(3)&&& \\
                                         &$V_{ph}$&0.0318(1)&&& \\
\hline
\end{tabular}
\end{center}
\end{table}

\begin{figure}
\begin{center}
\includegraphics[width=8cm]{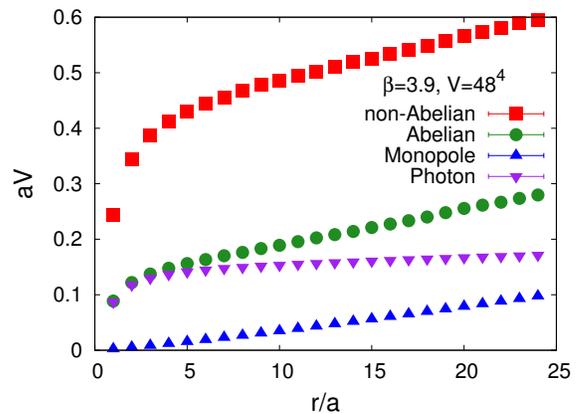}
\end{center}
\caption{The potential between quark and antiquark in the MCG. Only the data at $\beta=3.9$ on the $48^4$ lattice is shown as an example. }\label{MD}
\end{figure}

\subsection{Abelian and monopole dominances}
First, we check whether Abelian and monopole dominances observed in the MAG are seen in other smooth gauges like the MCG or not. We evaluate the potential from Abelian Wilson loops and their monopole contributions. Now, we take into account only a simple Abelian Wilson loop, say, of size $I \times J$. Then such an Abelian Wilson loop operator is expressed as
\begin{align}
W^a_{A}=\mathrm{exp}\{i\sum J_{\mu}(s) \theta_{\mu}^a(s) \},  
\end{align}
where $J_{\mu}(s)$ is an external current taking $\pm1$ along the Wilson loop. Since $J_{\mu}(s)$ is conserved, it is rewritten for such a simple Wilson loop in terms of an antisymmetric variable $M_{\mu\nu}$ as $J_{\nu}=\partial'M_{\mu\nu}(s)$ with a forward (backward) difference $\partial_{\nu}.(\partial'_{\nu})$. Note that $M_{\mu\nu}(s)$ take $\pm1$ on a surface with the Wilson loop boundary. Although we can choose any surface type, we adopt a minimal flat surface here. We get 
\begin{eqnarray}
 W^a_{A}=\mathrm{exp}\{-\frac{i}{2}\sum M_{\mu\nu}(s) \theta^{a}_{\mu\nu}(s)\}.
\end{eqnarray}
We investigate the monopole contribution to the static potential in order to examine the role of monopoles for confinement. The monopole part of the Abelian Wilson loop operator is extracted as follows \cite{Shiba:1994ab}. Using the lattice Coulomb propagator $D(s-s')$, which satisfies $\partial_{\nu} \partial'_{\nu} D(s-s') = -\delta_{ss'}$, we get
\begin{eqnarray}
W^a_{A}&=& W^a_{mon} W^a_{ph} , \\
W^a_{mon}&=&\mathrm{exp} \{2 \pi i \sum k^a_{\beta}(s) \notag \\
&\times& D(s-s^{'}) \frac{1}{2} \epsilon_{\alpha \beta \rho \sigma } \partial_{\alpha} M_{\rho \sigma}(s^{'}) \} , \\
W^a_{ph}&=&\mathrm{exp}\{-i \sum \partial^{'}_{\mu} \bar{\theta}^a_{\mu \nu}(s) D(s-s^{'}) J_{\nu}(s^{'}) \} .
\end{eqnarray}
We then compute the static potential from the Abelian Wilson loops and the monopole Wilson loops in the MCG and MAU1 on the $48^4$ lattices at $\beta =3.0, 3.2, 3.5, 3.9 $ and in the above four smooth gauges on $24^4$ at $\beta=3.5$. We fit the potential to the usual functional form
  \begin{align}
    V_{fit}(r) = \sigma r -c/r + \mu , 
  \end{align}
where $\sigma$ denotes the string tension, c the Coulombic coefficient, and $\mu$ the constant. The static potential in the MCG is shown in Fig.\ref{MD}. The results of the string tensions in the above four smooth gauges on the $24^4$ lattice are shown in Table \ref{fourGauge-sigma} and on $48^4$ in the MCG are summarized for various $\beta$ in Table \ref{MCG-sigma}. Here $V_{NA}$, $V_{A}$, $V_{mon}$ and $V_{ph}$ mean potentials from non-Abelian, Abelian, monopole, and photon Wilson loop, respectively. And $\sigma_{NA}$, $\sigma_A$, $\sigma_{mon}$ and $\sigma_{ph}$ are non-Abelian, Abelian, monopole, and photon string tensions. Fairly good results of Abelian and monopole dominances are obtained also in the MCG in comparison with those in the MAG. Both ratios $\sigma_{A}/\sigma_{NA}$ and $\sigma_{mon}/\sigma_{NA}$ approach more to one as the coupling constant $\beta$ becomes larger as expected from the previous data \cite{Suzuki:2007jp,Suzuki:2009xy}.  
\begin{table}[h] 
\caption{The string tension from the $n$ blocked monopole current in the MCG and the MAG. FR means the fitting range.}\label{block_sigma}
\begin{center}
\begin{tabular}{c|cccccc} \hline
&\ n\ &$\beta$ & $\sqrt{\sigma a^2}$ & FR(r/a)&$\chi^2/N_{d.o.f}$&$\sigma_{mon}/\sigma_{NA}$ \\ \hline
       &\ 1\ &3.0&$0.2910(1)$ &5-16&0.482225&0.7807\\
MCG&\ 2\ &3.4&0.296(2)&3-11&0.325642 &0.815\\
      &\ 3\ &3.6&0.330(2)&2-8& 0.25499 &0.842\\
\hline 
    &\ 1\ &3.0&$0.3026(1)$ &2-20&0.995868&0.8119\\
MAG&\ 2\ &3.4&0.304(1)&2-11&0.919558 &0.836\\
     &\ 3\ &3.6&0.328(3)&2-8& 0.827499 &0.837\\
\hline
\end{tabular}
\end{center}
\end{table}
\begin{figure}
\begin{center}
\includegraphics[width=8cm]{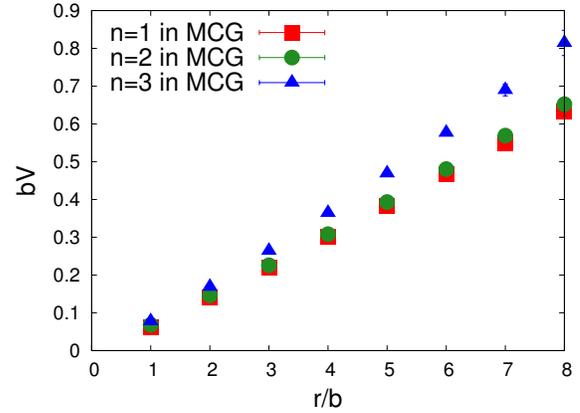}
\end{center}
\caption{The static-quark potentials from monopole Wilson loops on a blocked reduced lattice with the spacing $b=na$ in the MCG. The data at $\beta=3.0$ is without the block-spin transformation. The data at $\beta = 3.4$ ($\beta=3.6$) are taken from $n=2$ ($n=3$) blocked monopoles.}\label{block}
\end{figure}

\subsection{Monopole dominance after block-spin transformations of monopoles}
 Considering the previous data \cite{Suzuki:2007jp,Suzuki:2009xy} showing perfect monopole dominance, insufficient monopole dominance obtained here after smooth gauge fixings suggests that there still remain lattice artifact monopoles. Here, let us consider a block-spin transformation with respect to lattice monopoles.  After the block-spin transformation of monopoles, we can study the monopole behaviors in the long-range regions near to the continuum limit. In Ref.\cite{Suzuki:2017lco}, the scaling behavior is seen when the monopole density is plotted versus the lattice spacing of the blocked lattices $b=na(\beta)$.  This result suggests the contribution of the monopole on the blocked lattice must be larger than that of the monopole on the original lattice.  We evaluate the monopole Wilson loop in the MCG and the MAG by using the monopole currents on the blocked $24^4$ ($16^4$) lattice after the $n=2$ ($n=3$) block-spin transformation of monopoles on the original $48^4$ lattice. Here the definition of the block-spin transformation of the monopole current is shown as 
\begin{align}
 k^{(n)}_{\mu}(s_n) = \sum_{i,j,l=0}^{n-1} k_{\mu}(ns + (n-1)\hat{\mu} + i\hat{\nu} +j\hat{\rho} + l\hat{\sigma}) .
\end{align} 
In the calculations of physics on a blocked lattice, it is important to adopt a corresponding improved operator measuring physics correctly as well as the effective monopole action on the blocked lattice \cite{Chernodub:2000}. But in the case of measuring the string tension, it is enough to consider flat Wilson loops on the blocked lattice as an improved operator. We evaluate the monopole contribution to the string tension at $\beta= 3.0$ on the original lattice $48^4$ and at $\beta=3.4$ ($\beta=3.6$) on the $n=2$ ($n=3$) blocked lattice $24^4$ ($16^4$). These $\beta$ points have  similar $b=na(\beta)$ values. As a result, the string tensions from the monopole Wilson loop on the blocked lattices in the MCG are larger than that on the original lattice as seen in Fig \ref{block}.   The string tensions from monopoles in the MAG and the MCG are summarized in Table \ref{block_sigma}. After $n=3$ blocking, the improvement in the MCG is bigger than that in MAG and the results in both gauges are almost the same. This is consistent with the results showing gauge independence obtained in previous work \cite{Suzuki:2017lco}. 
\begin{figure}
\begin{center}
\includegraphics[width=7cm]{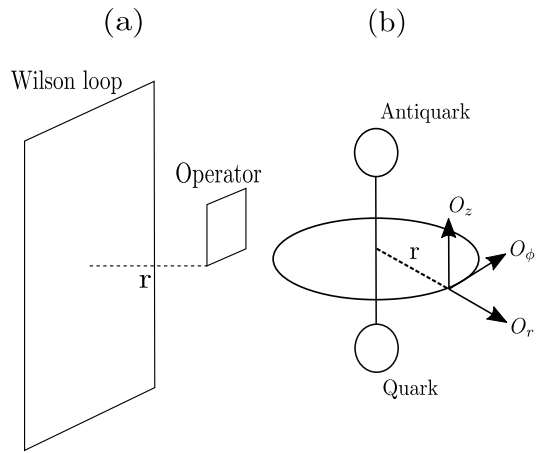}
\end{center}
\caption{(a) is the schematic figure of the disconnect correlation between an Abelian Wilson loop and an Abelian operator. (b) is the definition of the cylindrical coordinate $(r,\phi,z)$ along the $q$-$\bar{q}$ axis. }\label{pic}
\end{figure}

\begin{figure}
\begin{center}
\includegraphics[width=8cm]{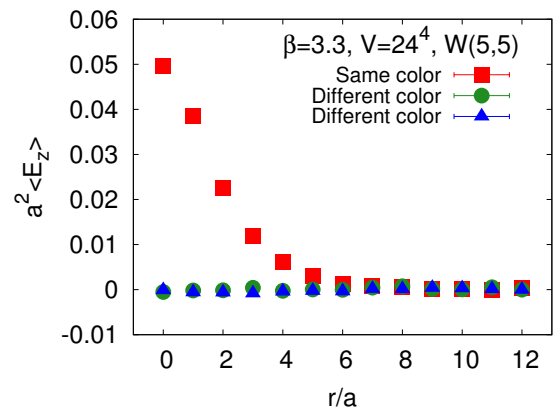}
\end{center}
\caption{The color distribution of electric fields $E_z$ on the $24^4$ lattice in the MCG. Only the $\beta=3.3$ case is plotted.}\label{E-color}
\end{figure}

\begin{figure}
\begin{center}
\includegraphics[width=8cm]{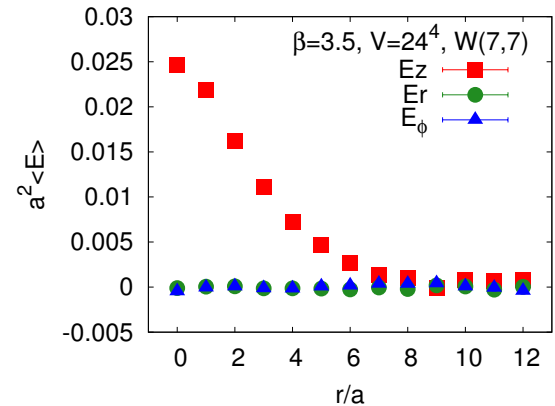}
\end{center}
\caption{The distribution of electric-field components $E_z$, $E_r$ and $E_{\phi}$ on the $24^4$ lattice in the MCG. Only the $\beta=3.5$ case is shown.}\label{E-zrphi}
\end{figure}
\begin{figure}
\begin{center}
\includegraphics[width=8cm]{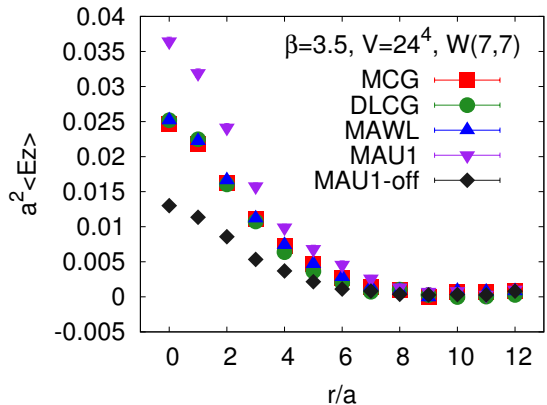}
\end{center}
\caption{The profiles of the color electric field $E_z$ at $\beta=3.5$ on the $24^4$ lattice for four smooth gauge fixings}\label{E}
\end{figure}
\begin{figure}
\begin{center}
\includegraphics[width=8cm]{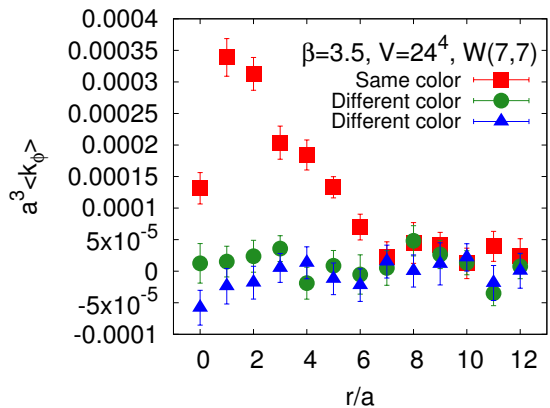}
\end{center}
\caption{The profiles of monopole current $k_{\phi}$ distributions at $\beta=3.5$ on the $24^4$ lattice in the MCG. There is no correlation between different colors.}\label{col}
\end{figure}
\begin{figure}
\begin{center}
\includegraphics[width=8cm]{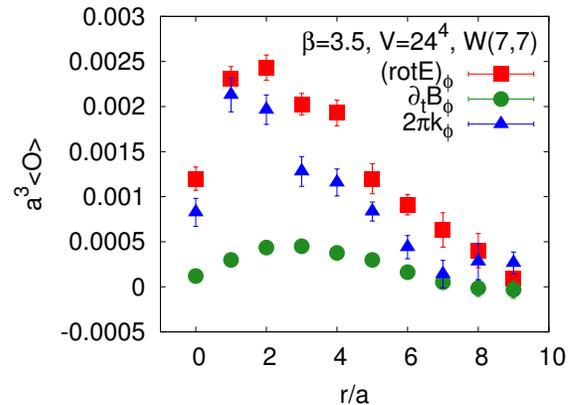}
\end{center}
\caption{The dual Amp$\grave{\mathrm{e}}$re's law at $\beta=3.5$ on the $24^4$ lattice in the MCG. }\label{dA}
\end{figure}


\subsection{The dual Meissner effect\label{DME}}
Next, we show the results with respect to the Abelian dual Meissner effect. It is necessary to measure the correlation functions between an Abelian Wilson loop and various Abelian operators having the same, or different colors. But in the previous research \cite{Suzuki:2009xy}, without any gauge fixing they could measure only the correlations between a non-Abelian Wilson loop and Abelian operators, which are connected by a Schwinger line, since the disconnected correlations  are too small to get a reliable result. The connected correlations, however, contain various contaminations, and it is desirable to measure original disconnected correlations between an Abelian Wilson loop and Abelian operators directly.
Therefore, we here adopt the above four gauge fixings smoothing the vacuum and evaluate such disconnected correlation functions:
\begin{align}
  \rho(W^a,O^b) = \frac{\braket{W^a O^b}-\braket{W^a}\braket{O^b}}{\braket{W^a}}, \label{rho}
 \end{align}
where $W^a$ is an Abelian Wilson loop, and $O^b$ is an Abelian operator. Here $a$ and $b$ denote color indices. A schematic figure and the definition of coordinates are depicted in Fig.\ref{pic}. 
In this simulation we adopt Wilson loops of W(R=3,T=3) at $\beta=3.0$, W(R=5,T=5) at $\beta=3.3$, and W(R=7,T=7) at $\beta = 3.5$ on the $24^4$ lattice. The physical $q-\bar{q}$ distances are almost equal to 0.48 fm for these Wilson loops. 
\subsubsection{Color electric field distributions}
To evaluate the profile of the Abelian color electric field, we calculate the correlation between an Abelian Wilson loop and an Abelian plaquette. In the naive continuum limit $a \rightarrow 0$, the correlation becomes $\braket{E_{i}}_{q\bar{q}} $. From now on, only the MCG is discussed among global color invariant gauges since the behaviors in the DLCG and the MAWL are much the same as those in the MCG.

The results are as follows:
\begin{enumerate}
  \item When we put a static quark-antiquark producing an adjoint color flux,  Abelian electric fields with the same color alone  exist around the quark pair as is naturally expected. It is shown in Fig.\ref{E-color}.
  \item The Abelian electric fields are squeezed actually. Figure \ref{E-zrphi} shows the electric-field components at the midpoint between the quark and the antiquark pair. The electric field runs parallel to the direction between the quark and the antiquark static sources.
  \item Figure \ref{E} shows $\braket{E_z}_{q\bar{q}}$ in four smooth gauge fixings. In the case of MAU1, the global color symmetry is broken. Hence, we evaluate both the diagonal component and the off-diagonal one separately. These data are fitted to a function
\begin{align}
f(r) = c_1 \mathrm{exp}(-\frac{r}{\lambda}) + c_0. \label{pene}
\end{align}
The parameter $\lambda$ corresponds to the penetration depth and the values for different gauge fixings are summarized in Table \ref{pene35}. The difference appears only with respect to the coefficient $c_1$ in the fitting function Eq.(\ref{pene}). These results show that there is little gauge dependence with respect to the behavior of the squeezing of the Abelian color electric field.  
\end{enumerate}
\begin{table}[h] 
\caption{The penetration length at $\beta = 3.5$}\label{pene35}
\begin{center}
\begin{tabular}{cccc} \hline
$$ & $\lambda$[fm]& $c_1$ & $ c_0 $  \\ \hline
MCG&0.189(16)&0.0330(12)&-0.00045(44) \\
\hline
DLCG&0.175(13)&0.0352(12)&-0.00067(36)\\
\hline
MAWL&0.189(16)&0.0336(13)&-0.00043(45)\\
\hline
MAU1&0.190(14)&0.0482(15)&-0.00065(53) \\
\hline
MAU1(off-diagonal)&0.175(17)&0.0175(8)&-0.000(2) \\
\hline
\end{tabular}
\end{center}
\end{table}
\subsection{The monopole-current distribution}
We then evaluate the monopole-current $k_{i}^b$ distributions around the static quark and antiquark pair defined by the relation
\begin{align}
  \rho(W^a,k_{i}^b) = \frac{\braket{W^a k_{i}^b}}{\braket{W^a}} ,\label{EqW-k}
 \end{align}
where $a$ and $b$ are color indices.
\subsubsection{The correlation between different color objects}
It is interesting to see the color correlation between the color of the static quark source and that of monopoles keeping the global color invariance. Here  we evaluate the correlations adopting the above three smooth gauges keeping the global color invariance at three different couplings $\beta=3.0, 3.3, 3.5$. The example of the $k_{\phi}$ distribution in the MCG case is shown in Fig.\ref{col}.   We find the peak of the signal of the monopole current(VNABI) slightly away from the flux-tube. There are no correlations between different colors.  This result means that an Abelian color electric field is squeezed by the same color monopole current alone. This is consistent with the Abelian confinement picture. 
\subsubsection{The dual Amp$\grave{e}$re's law}
To see what squeezes the color-electric field, we investigate the  dual Amp$\grave{\mathrm{e}}$re's law derived from the definition of the monopole current
   \begin{align}
 (\mathrm{rot}E^{a})_\phi = \partial_{t}B^{a}_{\phi} + 2\pi k^{a}_{\phi} , \label{dualA}
\end{align}
where index $a$ is a color index with $a=1,2,3$. 
We confirm the dual Amp$\grave{\mathrm{e}}$re's law holds in four smooth gauge fixings. As a typical global-color invariant gauge, we show graphs for the MCG alone in Fig.\ref{dA}. The electric field is squeezed mainly due to Abelian monopole currents as obtained in the MAG \cite{Matsubara:1994} and in the work without any gauge fixing \cite{Suzuki:2009xy}.  In the case of MAU1, the global color invariance is broken. With respect to the diagonal component in MAU1 gauge, Abelian monopole currents are shown to squeeze the electric field \cite{Matsubara:1994}. But the behavior of the off-diagonal component looks different. In this case, the Abelian color magnetic displacement current $\partial_t B$ seems to play the role of squeezing the off-diagonal electric field instead of the Abelian monopole current like in the Landau gauge \cite{Suzuki:2005}. But in the MAU1 case, it is only apparent, since even the off-diagonal components contain monopoles if lattice artifacts are deleted enough as studied in Ref.\cite{Suzuki:2017lco}, whereas in the Landau gauge, lattice monopoles \`{a} la Degrand-Toussaint\cite{DeGrand:1980eq} do not exist.

\begin{figure}
\begin{center}
\includegraphics[width=8cm]{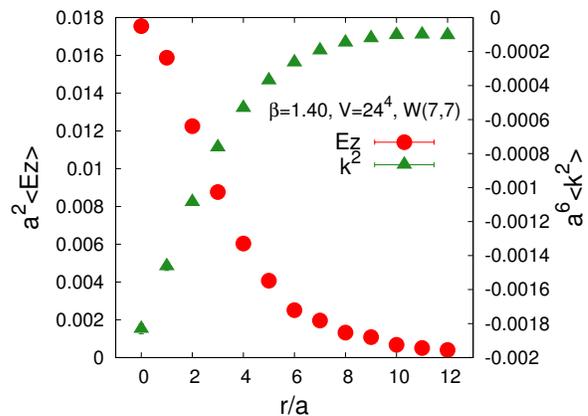}
\end{center}
\caption{The behaviors of the electric field squeezing and the monopole density at $\beta = 1.40$ on the $24^4$ lattice in the MCG.}\label{E-K2}
\end{figure}

\begin{table}[h] 
\caption{The GL parameter in the MCG}\label{GL}
\begin{center}
\begin{tabular}{|c|c|c|c|} \hline
 \multicolumn{1}{|c|}{$\beta$} &  \multicolumn{1}{|c|}{$\lambda$[fm]}&  \multicolumn{1}{|c|}{$\xi/\sqrt{2}$ [fm]}  &  \multicolumn{1}{|c|}{$\sqrt{2}\kappa$} \\ \hline
 1.10&0.124(2)&0.121(2)&1.022(1)  \\
 1.28&0.105(4)&0.087(3)&1.208(3) \\
 1.40&0.144(8)&0.128(5)&1.13(1) \\
\hline
\end{tabular}
\end{center}
\end{table}
\subsection{The vacuum type in the MCG}
Finally, we evaluate the Ginzburg-Landau (GL) parameter, which characterizes the type of the (dual) superconducting vacuum. In the previous result \cite{Suzuki:2009xy}, they found that the vacuum type is near the border between the type 1 and type 2 dual superconductors by using the SU(2) Iwasaki action without gauge fixing. The SU(2) Iwasaki action is adopted also to make a comparison with the previous result \cite{Suzuki:2009xy}. The Iwasaki improved action is essentially the same as (\ref{tad}) except the mixing parameter. Here, we  evaluate the GL parameter in the case of a smooth MCG.  The lattices adopted are $24^4$ for $\beta=1.10, 1.28, 1.40$. \\
\ \ The GL parameter is the ratio of the penetration length and the coherence length. The penetration length is measured as done previously, in the tadpole-improved action (\ref{tad}). To evaluate the coherence length, we evaluate the correlation between the squared monopole density $\sum_{\mu}k_{\mu}(s)k_{\mu}(s)$ and the Abelian Wilson loop by using the disconnected correlation function. The typical data is shown in Fig.\ref{E-K2}. We fit the profile of $\braket{\sum_{\mu}k_{\mu}(s)k_{\mu}(s)}_{q\bar{q}}$ to the function
\begin{align}
g(r) = c^{\prime}_1\mathrm{exp}(-\frac{\sqrt{2}r}{\xi}) + c^{\prime}_0 ,
\end{align}
where the parameter $\xi$ corresponds to the coherence length. The number of gauge configurations is $N_{conf}=1000$ to get the signal of the correlation. 
We show the result of the GL parameter $\kappa = \lambda/\xi$ in the Table \ref{GL}. The GL parameter in MCG is close to the value of the previous result \cite{Suzuki:2009xy}. These  show that the vacuum after the smooth MCG captures the essential property of the vacuum in SU(2) gauge theory with a reasonable number of field configurations as opposed to the case of no gauge fixing. 
\section{Conclusion}
 In conclusion, we have studied monopole dominance and the dual Meissner effect in three smooth gauge fixings which preserve  global color symmetry as well as in the MAG. The summary is depicted as follows:
  \begin{enumerate}
   \item The string tension of the static potential is reproduced fairly well by the monopole contribution. When the string tension is evaluated after the block-spin transformation of monopoles, the monopole dominance is improved. The value of the string tension in the MCG and the MAG become about the same on the blocked lattice. These results suggest that perfect monopole dominance and gauge independence are realized in the continuum limit.
   \item In the study of the dual Meissner effect due to Abelian-like monopoles, the electric field having a color is squeezed by the corresponding colored monopoles alone, as predicted by the Abelian picture of confinement.  We find the scaling behavior of the dual Meissner effect in four gauge fixings. 
   \item The vacuum type is determined to be at the border between type 1 and type 2 in SU(2) gauge theory with the smooth MCG gauge. This is consistent with the previous data without gauge-fixing \cite{Suzuki:2009xy}.
\item The Abelian monopoles here correspond to VNABI in the continuum limit which are gauge variant. Hence, we have to adopt any method extracting the continuum gauge-invariant part on the lattice. One way is to adopt a very large number of vacuum ensembles for an average as adopted in Ref.\cite{Suzuki:2009xy}.  Another method is to adopt a vacuum ensemble which is smooth enough  to reduce the lattice artifacts as much as possible. In this sense, adopting a special gauge is important. The MAG is the smoothest gauge known so far. Here, we show that the MCG is also a good gauge which can reproduce roughly the essential monopole properties of the continuum SU(2) QCD with a reasonable number of field configurations similarly as in the MAG. Moreover, contrary to the MAG, the MCG has the advantage of preserving the global color invariance and is so very interesting. To study the correlation between the Abelian monopoles and the center vortex in the MCG may also be interesting, since the MCG was first discussed in the framework of the center vortex model \cite{DelDebbio:1996mh,DelDebbio:1998uu}.
\item
Since the Abelian-like monopoles studied in this work and the previous works\cite{Suzuki:2017lco,Suzuki:2017zdh} have a gauge-invariant continuum limit, it is very important to study what quantity corresponds to the limit in the framework of continuum QCD. 
\item In the Abelian projection scenario of color confinement proposed by 't Hooft \cite{tHooft:1981ht}, Abelian monopoles appear as topological objects corresponding to the homotopy group by adopting a partial gauge fixing. There the singularity leading to Abelian monopoles comes from the partial gauge fixing. In our scenario, VNABI comes from a line singularity 
possibly
existing in original gauge fields. If this scenario is correct, we have to deal with a field theory composed of an operator having such a singularity. Such a singular operator is not considered in the framework of usual axiomatic field theory. It is interesting to extend a mathematical framework to accommodate such a singular operator. 
 \end{enumerate} 
 
 \section{ACKNOWLEDGMENTS}
The numerical simulations of this work were done using HPC and NEC SX-ACE computer at RCNP of Osaka University and partially by NEC SX-Aurora computer at KEK. The authors would like to thank RCNP and KEK for their support through computer facilities. T.S. was finacially supported by JSPS KAKENHI Grant No. JP19K03848. A.H. was financially supported by the Sasakawa Scientific Research Grant from
The Japan Science Society.

\end{document}